\documentclass[prl,twoside,twocolumn,10pt,floatfix,showpacs,citeautoscript,superscriptaddress]{revtex4-2}

\usepackage{amsmath}
\usepackage{amssymb}
\usepackage{booktabs}
\usepackage{comment}
\usepackage{glossaries}
\usepackage{graphicx}
\usepackage[version=4]{mhchem}
\usepackage{siunitx}
\usepackage{tabularx}
\usepackage[dvipsnames]{xcolor}

\usepackage[colorlinks=true]{hyperref}
\hypersetup{
    pdffitwindow=false,
    pdfstartview={FitH},
    pdfnewwindow=true,
    colorlinks=true,
    linkcolor=NavyBlue,
    citecolor=NavyBlue,
    filecolor=NavyBlue,
    urlcolor=NavyBlue,
}

\hypersetup{
    pdfauthor={Erik Fransson, Petter Rosander, Fredrik Eriksson, J. Magnus Rahm, Terumasa Tadano, and Paul Erhart},
    pdftitle={Limits of the phonon quasi-particle picture at the cubic-to-tetragonal phase transition in halide perovskites}
}

\graphicspath{{figures/}}

\setacronymstyle{long-short}
\newacronym{acf}{ACF}{autocorrelation function}
\newacronym{dft}{DFT}{density functional theory}
\newacronym{dho}{DHO}{damped harmonic oscillator}
\newacronym{ehm}{EHM}{effective harmonic model}
\newacronym{fc}{FC}{force constant}
\newacronym{fcp}{FCP}{force constant potential}
\newacronym{gpu}{GPU}{graphical processing unit}
\newacronym{md}{MD}{molecular dynamics}
\newacronym{mlp}{MLP}{machine-learned potential}
\newacronym{nep}{NEP}{neuroevolution potential}
\newacronym{pes}{PES}{potential energy surface}
\newacronym{rfe}{RFE}{recursive feature elimination}
\newacronym{rmse}{RMSE}{root-mean-square error}
\newacronym{scan}{SCAN}{strongly constrained and appropriately normed}
\newacronym{sscha}{SSCHA}{stochastic self-consistent harmonic approximation}
\newacronym{scp}{SCP}{self-consistent phonon}
\newacronym{sed}{SED}{spectral energy density}
\newacronym{soap}{SOAP}{smooth overlap of atomic positions}

\DeclareSIUnit\angstrom{\text {Å}}
\renewcommand{\vec}[1]{\ensuremath\boldsymbol{#1}}
\renewcommand{\epsilon}[0]{\varepsilon}

\global\let\oldnewlabel\newlabel
\gdef\newlabel#1#2{\newlabelxx{#1}#2}
\gdef\newlabelxx#1#2#3#4#5#6{\oldnewlabel{#1}{{#2}{#3}}}
\let\newlabel\oldnewlabel

\newcommand{\addchalmers}{
    Department of Physics,
    Chalmers University of Technology,
    SE-41296, Gothenburg, Sweden
}
\newcommand{\addnims}{
    Research Center for Magnetic and Spintronic Materials,
    National Institute for Materials Science (NIMS),
    1-2-1 Sengen, Tsukuba, Ibaraki 305-0047, Japan
}

\begin{document}

\title{
    Limits of the phonon quasi-particle picture\texorpdfstring{\\}{}
    at the cubic-to-tetragonal phase transition in halide perovskites
}

\author{Erik Fransson}
\author{Petter Rosander}
\author{Fredrik Eriksson}
\author{J. Magnus Rahm}
\affiliation{\addchalmers}
\author{Terumasa Tadano}
\affiliation{\addnims}
\author{Paul Erhart}
\email{erhart@chalmers.se}
\affiliation{\addchalmers}

\begin{abstract}
The soft modes associated with continuous-order phase transitions are associated with strong anharmonicity.
This leads to the overdamped limit where the phonon quasi-particle picture can breakdown.
However, this limit is commonly restricted to a narrow temperature range, making it difficult to observe its signature feature, namely the breakdown of the inverse relationship between the relaxation time and damping.
Here we present a physically intuitive picture based on the relaxation times of the mode coordinate and its conjugate momentum, which at the instability approach infinity and the inverse damping factor, respectively.
We demonstrate this behavior for the cubic-to-tetragonal phase transition of the inorganic halide perovskite \ce{CsPbBr3} via molecular dynamics, and show that the overdamped region extends almost \SI{200}{\kelvin} above the transition temperature.
Further, we investigate how the dynamics of these soft phonon modes change when crossing the phase transition.
\end{abstract}

\maketitle

% ===============
%% Introduction
%% ==============
\section{Introduction}
The vibrational properties of solids are pivotal for many physical phenomena, including but not limited to phase stability and thermal conduction.
In crystalline solids, the vibrational spectrum is commonly described in terms of phonons as quasi-particle representations of the lattice vibrations.
The phonon frequency $\omega_0$ is typically much larger than the damping $\Gamma$, and the phonon relaxation time $\tau = 2/\Gamma$ is thus much longer than the oscillation period, such that the quasi-particle picture is well motivated \cite{Zim60, SunAll2010, SunSheAll2010, ZhaSunWen2014, Lv2016, Isaeva2019}.
In this so-called underdamped limit, the relaxation time \emph{decreases} as the damping $\Gamma$ increases.

By comparison, there are far fewer cases when phonon modes become overdamped, i.e., $\omega_0 \tau < 1$ \cite{Silverman1974, SchSto1978}.
This can occur either due to large damping or for very soft modes, usually in the immediate vicinity of a phase transition, as for example in the case of body-centered cubic Ti \cite{PetHeiTra91, FraErh20, FraSlaErhWah2021}, rotationally disordered 2D materials \cite{Kim2021}, in ferroelectrics such as \ce{BaTiO3} \cite{Nakamura1975, Nakamura1992, Dove1997, Ehsan2021, Verdi2022} or in halide perovskites \cite{Songvilay2019, Lanigan2021}.
In the overdamped limit, the relaxation time \emph{increases} with increasing damping $\Gamma$, which calls into question the picture of a well-defined phonon mode with a frequency and relaxation time.
Overdamped phonon dynamics is, however, usually limited to a rather narrow temperature window and under these circumstances the inversion of the relation between relaxation time and damping cannot be readily observed.
Here, we demonstrate that the soft phonons modes associated with the phase transitions in the prototypical halide perovskite \ce{CsPbBr3} are, however, outstanding manifestations of this exact behavior as the overdamped region extends almost \SI{200}{\kelvin} above the tetragonal-cubic phase transition.

\begin{figure*}
\centering
\includegraphics{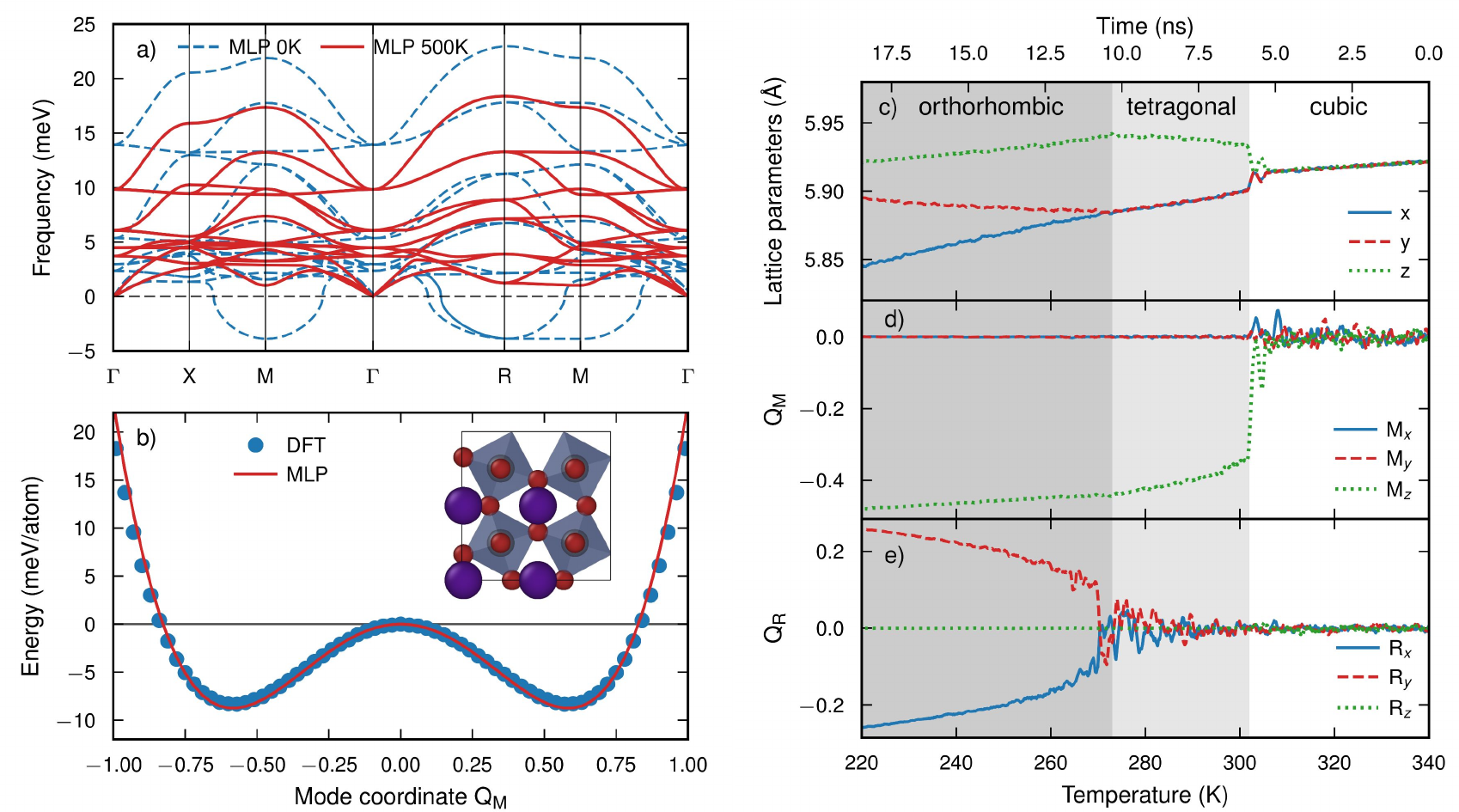}
\caption{
    \textbf{Phonons and phase transitions in \ce{CsPbBr3}.}
    a) Phonon dispersion for the cubic phases of \ce{CsPbBr3} obtained using the machine learned potential (MLP) in the harmonic approximation, \SI{0}{\kelvin}, (dashed lines) and from an effective harmonic model (EHM) at \SI{500}{\kelvin} (solid lines).
    b) Potential energy landscape along the unstable M-tilt mode calculated with \gls{mlp} and density functional theory (DFT).
    The inset shows the \ce{CsPbBr3} crystal structure (Cs purple, Pb gray, Br red) in the energy minima, for which the \ce{PbBr6} octahedra have been tilted in-phase (visualization made with \textsc{ovito} \cite{Stukowski2010}).
    c, d, e) Lattice parameters and mode coordinates obtained from a cooling run based on the isothermal-isobaric ($NpT$) ensemble with phase transitions at approximately \SI{300}{\kelvin} and \SI{275}{\kelvin}.
    In c, d, e) solid, dashed and dotted lines refer to the Cartesian directions x, y and z respectively.
    }
\label{fig:phonons_and_phase_transitions}
\end{figure*}

Halide perovskites are promising materials for photovoltaic and optoelectronic applications.
Specifically, \ce{CsPbBr3} has received a lot of attention in recent years \cite{Stoumpos2013}.
With increasing temperature it undergoes phase transitions from an orthorhombic (Pnma) to a tetragonal (P4/mbm) and eventually a cubic phase (Pm$\bar{3}$m) \cite{Hirotsu1974, Sharma1991, Rodov2003, Lopez2020, Malyshkin2020}.
These phase transitions are connected to specific phonon modes and arise due to tilting of the \ce{PbBr6} octahedra, corresponding to phonon modes at the R and M points (\autoref{fig:phonons_and_phase_transitions}a) \cite{Huang2014, daSilva2015, Yang2017, Klarbring2018, Yang2020}.
Experimentally, these modes have been shown to exhibit overdamped characteristics in the vicinity of the phase transitions \cite{Songvilay2019, Lanigan2021, Cohen2022}.
The phase transitions have also been studied from first-principles and via \gls{md} simulations, see, e.g., Refs.~\onlinecite{Klarbring2019, ZhuEgger2022, Lahnsteiner2022, TadWis2022}.

Here, we reveal the dynamics of the octahedral tilt modes in \ce{CsPbBr3} over a wide temperature range via \gls{md} simulations based on a \gls{mlp} that achieves close to \gls{dft} accuracy (\autoref{snote:model_validation}, \autoref{sfig:model_validation}) \cite{FanZenZha21, FanWanYin22}.
To obtain access to mode specific dynamics we project the \gls{md} trajectories onto different normal modes that are associated with the phase transitions in this material.
As shown below, this requires both large large systems (comprising at least several \num{10000} atoms) and sufficiently long times scales ($\sim\,\SI{50}{\nano\second}$ to \SI{100}{\nano\second}) in order to achieve converged results (see \autoref{snote:mode_projections}, \autoref{sfig:acf_size_convergence} and \autoref{sfig:acf_time_convergence}).
The \gls{dft} data and the \gls{mlp} models are provided as a Zenodo dataset \cite{zenodo_dataset}.

Reference data for the construction of the \gls{mlp} was generated by \gls{dft} calculations \cite{KreHaf93, Blo94, KreFur96} using the \gls{scan} exchange-correlation functional \cite{SunRuzPer15} (\autoref{snote:dft}).
Simulations and atomic structures were handled via the \textsc{ase} \cite{Larsen2017} and \textsc{calorine} packages \cite{calorine}.
The obtained phonon frequencies and relaxation times with the \gls{mlp} are in good agreement with experimental work for multiple phonon modes (see \autoref{sfig:dispersion_G2X_G2M}).
In addition, we consider several different \gls{scp} renormalization methods \cite{Esfarjani2020} as well as \glspl{ehm} \cite{Kong2009, Kong2011, And12, HelSteAbr13} using the \textsc{hiphive} \cite{EriFraErh19}, \textsc{alamode} \cite{TadGohTsu14}, and \textsc{sscha} packages \cite{MonBiaChe21}.

\begin{figure}
\centering
\includegraphics{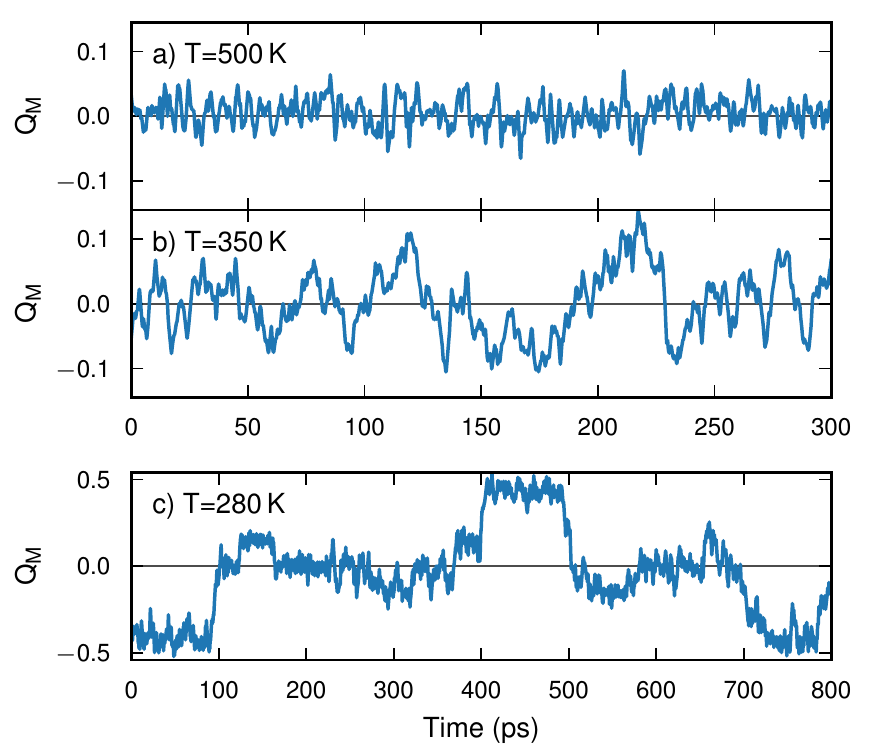}
\caption{
    \textbf{Phonon mode coordinates.}
    Mode coordinate $Q(t)$ for the M-tilt mode, a) at \SI{500}{\kelvin} (well above $T_{c\leftrightarrow t}$), b) at \SI{350}{\kelvin} (close to $T_{c\leftrightarrow t}$), and c) at \SI{280}{\kelvin} (below above $T_{c\leftrightarrow t}$).
    The M-tilt mode is three-fold degenerate ($x$, $y$, $z$) but here only $M_z$ show, and note that for \SI{280}{\kelvin} the system switches the tilt axis at irregular intervals.
}
\label{fig:mode_coordinates}
\end{figure}

% ===============
%% Results
%% ==============
\section{Results and discussion}

% tilt-modes and phase transitions
\textbf{Tilt-modes and phase transitions.}
The phase transitions in \ce{CsPbBr3}, and similarly in many other perovskites, are driven by modes that correspond to the tilting of the \ce{PbBr6} octahedra.
These modes are located at the M (in-phase tilting) and R-points (out-of-phase tilting) in the phonon dispersion for the cubic structure (\autoref{fig:phonons_and_phase_transitions}a).
They are three-fold degenerate corresponding to tilting around the three Cartesian directions.
These tilt-modes exhibit a double-well \gls{pes}, which the \gls{mlp} reproduces  perfectly compared to \gls{dft} (\autoref{fig:phonons_and_phase_transitions}b).

The \gls{mlp} predicts temperatures of \SI{300}{\kelvin} and \SI{275}{\kelvin} for the cubic-tetragonal, $T_{c\leftrightarrow t}$, and tetragonal-orthorhombic, $T_{t\leftrightarrow o}$, transitions respectively (\autoref{fig:phonons_and_phase_transitions}c).
This is lower than the experimental values of \SI{400}{\kelvin} and \SI{360}{\kelvin} \cite{Sharma1991, Rodov2003, Stoumpos2013, Malyshkin2020}, a discrepancy that can be primarily attributed to the underlying exchange-correlation functional \cite{FraWikErh2023}.

The mode coordinates of the tilt modes are useful order parameters for analyzing the phase transitions (\autoref{fig:phonons_and_phase_transitions}d,e).
At \SI{300}{\kelvin} the system transitions from the cubic to the tetragonal phase as seen in both the lattice parameters and in the freezing in of one of the three M-tilt modes (M$_z$).
For the tetragonal phase two R-modes (R$_x$ and R$_y$) start to show larger fluctuations, and at \SI{265}{\kelvin} the system transitions to the orthorhombic phase.
Here, we also note the slight difference in character between these two phase transitions.
For the cubic-tetragonal transition the order parameter ($Q_\text{M}$) and lattice parameter change sharply at the transition temperature $T_{c\leftrightarrow t}$ (closer in character to a first-order transition), whereas for the tetragonal-orthorhombic transition the order parameter and lattice parameter change more gradually around $T_{t\leftrightarrow o}$ (exhibiting a continuous character) in agreement with experimental observations of the transition character \cite{Hirotsu1974, Stoumpos2013}.
We note here that the mode coordinate is a global order parameter for the system.
In the cubic phase even though the mode coordinate is on average zero, there still exists a strong local correlation between the neighboring octahedra.
This connects to previous work on perovskites regarding the local atomic structure deviating from the cubic structure while globally still appearing cubic \cite{Fabini2016, Bertolotti2017, Wiktor2017, Yang2017, Klarbring2018, Levin2021, Malavasi2022}.

%% mode coordinate dynamics at high T, close to Tc and below Tc
\textbf{Mode coordinate dynamics.}
The mode coordinates exhibit interesting dynamical behavior already in the cubic phase far above the transition to the tetragonal phase, which can be conveniently observed in the time domain, \autoref{fig:mode_coordinates}a, b).
At \SI{500}{\kelvin} regular (phonon) oscillator behavior is observed, whereas for \SI{350}{\kelvin} (closer but still above $T_\text{C}$) a slower dynamic component becomes evident.
Finally, at \SI{280}{\kelvin} and thus below the phase transition, one observes the common oscillatory motion superimposed on a long timescale hopping motion between the two minima, corresponding to the (degenerate) tetragonal phase (\autoref{fig:phonons_and_phase_transitions}b).
We note here that the hopping frequency depends strongly on system size, and is thus not a good observable on its own.

\begin{figure}
\centering
\includegraphics{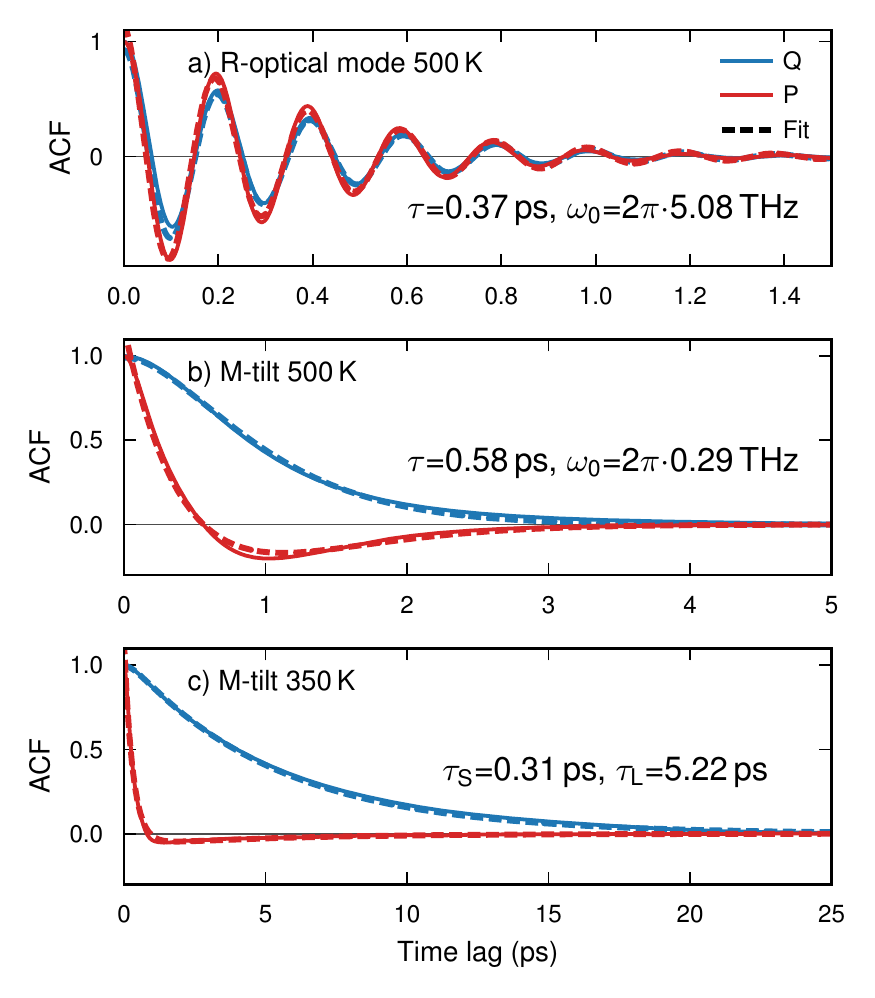}
\caption{
    \textbf{Phonon mode auto-correlation functions (ACFs).}
    ACFs (solid lines) of a) the highest optical mode at the R-point as well as the M-tilt mode at b) \SI{500}{\kelvin} and c) \SI{350}{\kelvin}, along with fits to a damped harmonic oscillator (DHO) model (dashed lines).
    Note the large difference in time-scale between b) and c).
}
\label{fig:mode_ACF}
\end{figure}

%% Autocorrelation and damped-harmonic oscillator
The mode coordinate can be analyzed by fitting the respective \glspl{acf} to a \gls{dho} model (\autoref{fig:mode_ACF}).
The \gls{acf} for a regular (underdamped) mode shows a clear oscillatory pattern as illustrated here by the highest optical mode at the R-point with a typical relaxation time of about \SI{0.37}{\pico\second}, which is longer than the mode period of about \SI{0.2}{ps} (\autoref{fig:mode_ACF}a).
The M-tilt mode at \SI{500}{\kelvin} has a similar damping but is much softer (yet still underdamped), and the \gls{acf} decays with a relaxation time of about \SI{0.58}{\pico\second} (\autoref{fig:mode_ACF}b).
At \SI{350}{\kelvin} (\autoref{fig:mode_ACF}c), however, the same mode is overdamped and in this case the \gls{dho} model becomes the sum of two exponential decays, see \autoref{eq:acf_Q_overdamped}, with relaxation  times $\tau_\text{L} = $\SI{5.22}{\pico\second} and $\tau_\text{S} = $\SI{0.31}{\pico\second}.
It is interesting to note that the decay time of the \gls{acf} at \SI{350}{\kelvin} is about ten-times longer than at \SI{500}{\kelvin}.
The \gls{dho} fits still match the data very well for both the underdamped and overdamped cases (see \autoref{sfig:acf_exp_splits} for how the two exponential decays behave for $Q$ and $P$ in the overdamped case).

When $\Gamma / \omega_0$ increases and the system becomes overdamped the dynamics of the modes is moving towards the diffusive Brownian motion regime.
For overdamped modes the relaxation time of the \gls{acf} increases as $\Gamma / \omega_0$ increases, opposite to the underdamped behavior.
While this is a well known feature of a simple one-dimensional \gls{dho}, here one observes this behavior for phonon modes in a complex atomistic system.
This phenomenon arises due to the \emph{free} energy landscape being very flat close to the transition, resembling a bathtub.
As a result of the high friction and weak restoring force, it therefore takes a long time for the \gls{dho} to move back and forth around zero (\autoref{fig:mode_coordinates}c; also see \autoref{sfig:Mtilt_powerspectra} for the power spectra) \cite{Volpe2013}.

\begin{figure}
\centering
\includegraphics{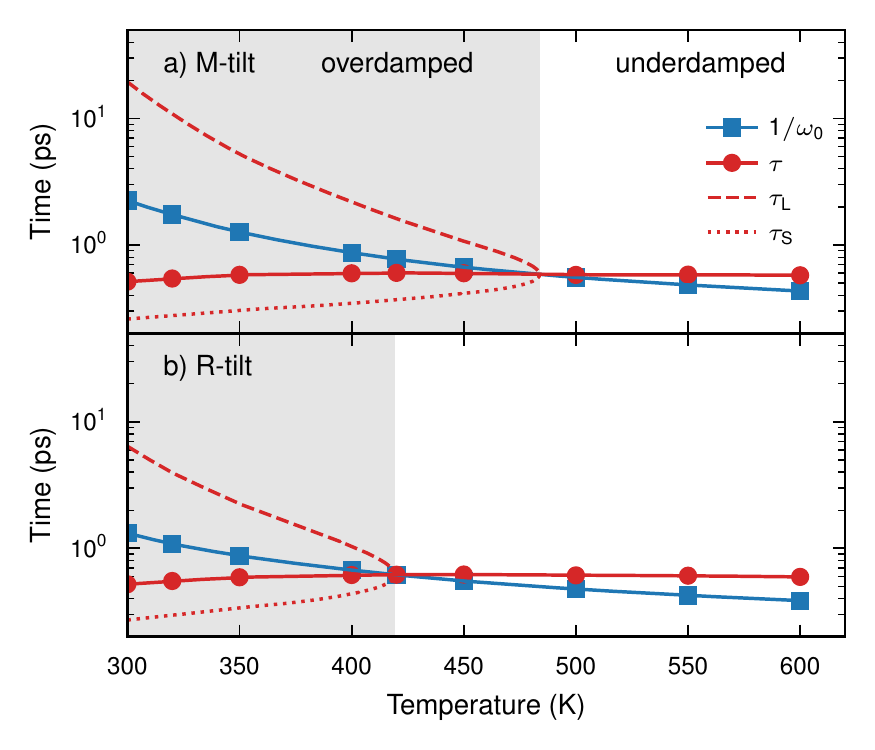}
\caption{
    \textbf{Phonon frequencies and relaxation times.}
    Frequencies and relaxation times obtained from auto-correlation functions of a) M-tilt and b) R-tilt modes as a function of temperature.
    The shaded region indicates the overdamped regime.
    Note that the M-tilt mode is overdamped already about \SI{200}{\kelvin} above the phase transition temperature.
    Here, markers are data points and lines are interpolations to guide the eye.
}
\label{fig:lifetimes_vs_temperature}
\end{figure}

%% Frequency and lifetimes vs temperature
\textbf{Frequency and relaxation times vs temperature.}
The frequencies and relaxation times of the M-tilt and R-tilt modes are summarized as a function of temperature in \autoref{fig:lifetimes_vs_temperature}.
The frequency $\omega_0$ softens significantly with decreasing temperature for both modes, whereas the relaxation time $\tau$ is more or less constant.
%We note that the frequency $\omega_0$ decreases almost linearly across the entire temperature range and does not follow $\omega_0^2 \sim (T-T_\text{C})$ as sometimes seen for displacive phase transitions \cite{Cochran1960, PytteFeder1969, Silverman1974}.
The softening of the frequency thus drives the modes to the overdamped limit with decreasing temperature.
The M-tilt and R-tilt modes only become underdamped above \SI{480}{\kelvin} and \SI{410}{\kelvin}, respectively, well above the transition temperature to the tetragonal phase at \SI{300}{\kelvin}.
This indicates that we expect the phonon quasi-particle for these modes to work better at high temperatures, which interestingly is the opposite behavior compared to most phonon modes which become more damped and anharmonic with increasing temperature.
At the cross-over from the underdamped to the overdamped regime, the two time scales $\tau_\text{S}$ and $\tau_\text{L}$ emerge.
When approaching $T_{c\leftrightarrow t}$  we see that $\tau_\text{L}$ increases exponentially, whereas  $\tau_\text{S} \to \tau/2$.

\begin{figure}
\centering
\includegraphics{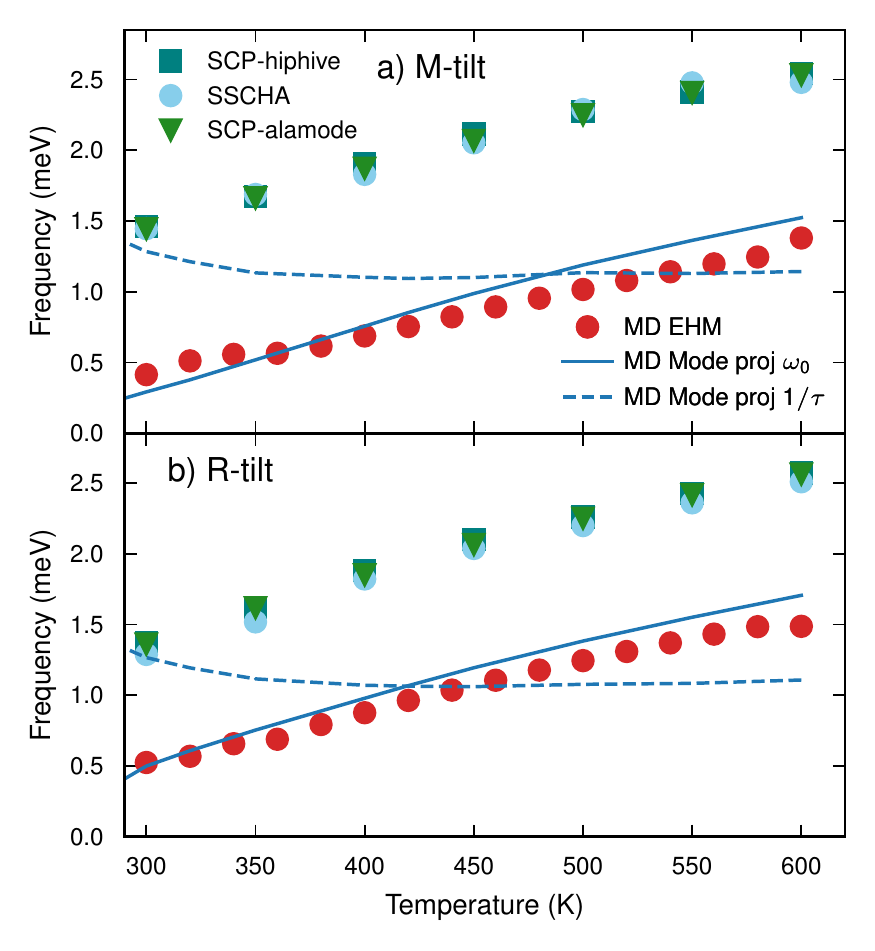}
\caption{
    \textbf{Phonon frequencies comparison with SCP.}
    Frequencies of a) M-tilt and b) R-tilt modes as a function of temperature from several self-consistent phonons (SCP) schemes, effective harmonic models (EHMs) based on molecular dynamics (MD), and damped harmonic oscillator (DHO) frequencies fitted to auto-correlation functions (ACFs).
}
\label{fig:frequencies_vs_temperature}
\end{figure}

%% SCP methods and EHMs
\textbf{Self-consistent phonons and effective harmonic models.}
Next, we analyze the representation of these strongly anharmonic modes by commonly used phonon renormalization techniques, specifically different \gls{scp} schemes and \glspl{ehm} (\autoref{fig:frequencies_vs_temperature}) (see \autoref{snote:scp} and \autoref{snote:ehm} for a more detailed description of the methods).
We construct 4:th order \glspl{fcp} at each temperature which is used as input to all \glspl{scp} methods, see \autoref{snote:fcp} and \autoref{sfig:scph_QM_CL_NEP_FCP} for more details.
There are several \glspl{scp} variants \cite{Esfarjani2020}.
In \gls{scp}-alamode the Green's function approach is employed as implemented in the \textsc{alamode} package \cite{TadGohTsu14}.
In \gls{sscha} the harmonic free energy is minimized using gradient methods, as implemented in the \textsc{sscha} package  \cite{MonBiaChe21}.
In \gls{scp}-hiphive second order force-constants are obtained by iterative fitting to forces from displacements sampled from the harmonic model and forces obtained from the \gls{mlp} as implemented in the \textsc{hiphive} package \cite{EriFraErh19}.
Here, we employ the ``bare'' \gls{scp} implementations in \textsc{alamode} and \textsc{sscha}.
We note, however, that there are computationally more demanding corrections for both methods \cite{Bianco2017, TadWis2022}, the analysis of which is, however, beyond the scope of the present work.
The \glspl{ehm} (in this field also referred to as temperature-dependent potentials) are constructed from fitting second-order force constants to displacement and force data obtained from \gls{md} simulations with the \gls{mlp} (see \autoref{snote:ehm} for details).

Here, we find very similar behavior for both M-tilt and R-tilt modes.
The three \gls{scp} methods (\gls{scp}-hiphive, \gls{sscha}, \gls{scp}-alamode) employed here are in great agreement with each other given the differences in theory and implementation between them.
The \gls{scp} frequencies systematically overestimate the frequency $\omega_0$ obtained from the \glspl{acf} by about \SI{1}{\milli\electronvolt} (see \autoref{snote:scp} for a more detailed description of the \gls{scp} methods).
The \glspl{ehm} constructed by fitting the  forces from \gls{md} trajectories, on the other hand, show good agreement with the mode projection results.
We note here that the trend for \glspl{scp} and \glspl{ehm} to over and underestimate frequencies, respectively, appears to hold for all modes in the system, which is in line with previous studies \cite{KorBelYan2018, MetKli2019, TadWis2022, Tolborg2022}.
However,  while \glspl{ehm} from \gls{md} yield a better frequency for the tilt modes compared to \gls{scp}, this is not in general true (see \autoref{sfig:R15_frequency} for details).

%% Tetragonal freqs, size dependency and modes becoming over-damped/under-damped. Characther of the two transitions. Fits Curie-Weiss law
\textbf{Behavior near phase transitions.}
Next, we look at how these modes behave as the system goes through the transition from the cubic to the tetragonal phase.
While in the cubic phase the three M (and R) modes (denoted with subscripts $x$, $y$, $z$ to indicate the Cartesian direction) are degenerate, the degeneracy is broken in the tetragonal phase and the $z$-direction becomes symmetrically distinct from the other two (\autoref{fig:phonons_and_phase_transitions}c,d).
Therefore, in order to distinguish these modes we will denote them by M$_{xy}$ and M$_{z}$ (analogously for R-modes) in the tetragonal phase.
In the tetragonal phase there exist multiple global minima for the M-mode coordinate, as can be seen in \autoref{fig:mode_coordinates}c where the system jumps between these minima.
To avoid capturing this (system-size dependent) hopping time in the \glspl{acf} we employ very large system sizes of up to \num{400000} atoms, for which the system remains in the same tetragonal orientation throughout the entire simulation.
Furthermore, we extrapolate the frequencies to the infinite system-size limit (\autoref{sfig:freq_size_conv}).

\begin{figure}
\centering
\includegraphics{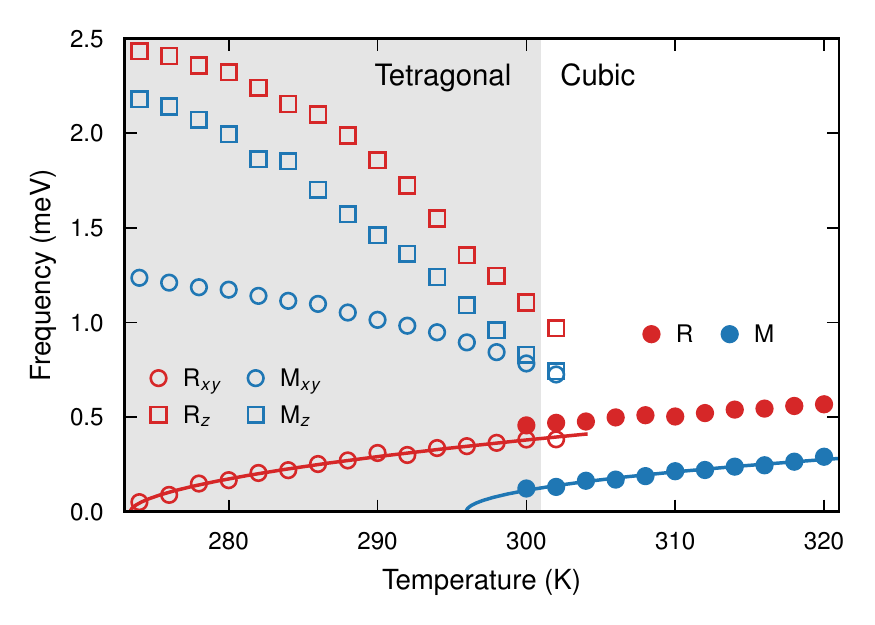}
\caption{
    \textbf{Phonon frequencies across the phase transition.}
    M and R-tilt mode frequencies from auto-correlation functions and damped harmonic oscillator fits as a function of temperature in the tetragonal and cubic phases.
    Here, filled symbols correspond to the frequencies in the cubic phase, and open symbols to the frequencies in the tetragonal phase.
    Solid lines show Curie-Weiss fits of the form, $\omega_0(T) \propto (T-T_\text{C})^p$.
    For the M-mode in the cubic phase we obtain $T_\text{C}=\SI{295}{\kelvin}$ (with fixed $p=0.5$), while for the R$_{xy}$ mode in the tetragonal phase we get $T_\text{C}=\SI{273}{\kelvin}$ and $p=0.55$.
    The cubic and tetragonal frequencies overlap in temperature, which is possible due to the first-order character of the transition and the fact that simulations are carried out in the microcanonical (NVE) ensemble which does not allow for the cell shape to change between cubic and tetragonal.
}
\label{fig:tetragonal_freqs}
\end{figure}

The resulting frequencies are shown in \autoref{fig:tetragonal_freqs}.
For the cubic-to-tetragonal transition the M-frequency does not go to zero at the transition temperature, which is in agreement with the character of the transition being first order, as observed experimentally \cite{Hirotsu1974, Rodov2003, Stoumpos2013}.
For the tetragonal-to-orthorhombic phase transition, on the other hand, the $R_{xy}$ frequency does go to zero at the transition temperature, in agreement with a continuous transition as observed experimentally \cite{Hirotsu1974}.
This leads to the long timescale in the \gls{dho} trending to infinity, $\tau_\text{L} \to \infty$, as the temperature approaches $T_\text{C}$.
Additionally, the $R_{xy}$ mode exhibits a strong size dependence close to the transition temperature (see \autoref{sfig:freq_size_conv}).

The Curie-Weiss law, $\omega_0(T) \propto (T-T_\text{C})^p$, provides very good fits for the temperature dependence of the modes driving the phase transitions.
For the tetragonal-to-orthorhombic transition the fitted critical temperature, $T_\text{C}=$\SI{273}{\kelvin}, agrees very well with the observed transition temperature of \SI{274+-1}{\kelvin}, which is consistent with this transition being a continuous transition \cite{Hirotsu1974}.
Furthermore, the fitted critical exponent of 0.55 is very close to the value of 1/2 suggested by Landau theory observed in many continuous phase transitions driven by soft modes \cite{Cochran1960, PytteFeder1969, Scott1974, Dove1997}.
The cubic-to-tetragonal transition has first-order character as evident from the finite frequency of the M mode at the transition temperature.
As a result, fitting both the critical temperature \emph{and} the critical exponent is ambiguous (due to the absence of data at temperatures for which the frequency goes to zero).
We therefore fix the critical exponent to 1/2, which yields a critical temperature of \SI{295}{\kelvin}, about \SI{7}{\kelvin} lower than the transition temperature.
Here, the critical temperature corresponds to the temperature at which the cubic phase becomes dynamically unstable, i.e., the point at which the free energy barrier between the two phases disappears.

The parameter $\tau$ remains fairly constant in the tetragonal phase across its entire temperature range for all four modes (\autoref{sfig:damping_tetragonal}).
Interestingly, once the M$_z$ mode freezes in (and the tetragonal phase is formed) both the M$_z$ and R$_z$ modes stiffen significantly with temperature.
This results in the R$_z$ mode becoming underdamped again with \emph{decreasing} temperature at around \SI{290}{\kelvin} and both M-modes approaching the underdamped limit as the system approaches the orthorhombic transition.

% ===============
%% Conclusions
%% ==============
\section{Conclusions}
We have carried out a detailed computational analysis of the dynamics in \ce{CsPbBr3}, focusing in particular on the tilt modes.
We observe overdamped modes for the cubic phase almost \SI{200}{\kelvin} above the cubic-to-tetragonal transition temperature.
These overdamped tilt-modes exhibit correlation on very long time scales ($\tau_\text{L}$) compared to the typical relaxation time ($\tau$) or period ($1/\omega_0$) of the mode.
This is in line with the dynamics of the modes transitioning toward Brownian motion due to the frequency approaching zero.
What we find here is that these modes can, however, still be mathematically well described as \glspl{dho}, which allows one to formally obtain a phonon frequency and relaxation time compliant with a quasi-particle picture.
A physically more intuitive description is, however, obtained if the \gls{dho} model is described by two relaxation times, which can be approximately associated with mode coordinate and momentum, respectively.
As a result of the soft character of these modes, the respective amplitudes can be large already at moderate temperatures.
This implies that even for relatively modest electron-phonon coupling strengths, these modes should have a notable impact on the optoelectronic properties of these materials \cite{Wiktor2017, Marronnier2018, Kirchartz2018, zhao2020polymorphous, wang2022accurate}.
A systematic investigation of these effects on a per-mode basis would be an interesting topic of further study.

In addition, we demonstrated that commonly used computational phonon renormalization methods agree very well with each other but without extensive correction schemes exhibit systematic errors in describing the frequencies of the anharmonic tilt modes considered here.
Understanding the single-point frequencies obtained from such methods and their relation to the full dynamical spectra is thus very important when, e.g., comparing to experimental measurements.

% ===============
%% Methods
%% ==============
\section{Methods}
To analyze phonon modes directly from \gls{md} simulations we employ phonon mode projection \cite{SunSheAll2010, CarTogTan2017, RohLiLuoHen2022}.
The \gls{md} simulations are carried out using the \textsc{gpumd} package \cite{FanWeiVie2017, FanWanYin22}.
For more details on the \gls{md} simulations see \autoref{snote:md}.
The atomic displacements $\vec{u}(t)$ and velocities $\vec{v}(t)$ can be projected on a mode $\lambda$, with the supercell eigenvector $\vec{e}_\lambda$ via
\begin{align*}
    Q_\lambda (t) = \vec{u}(t) \cdot \vec{e}_\lambda \quad\text{and}\quad
    P_\lambda (t) = \vec{v}(t) \cdot \vec{e}_\lambda.
\end{align*}
Here, the phonon supercell eigenvector of the tilt modes are obtained with \textsc{phonopy} \cite{TogTan15}, and symmetrized such that each of the three degenerate modes corresponds to tilting around the x, y and z direction respectively.
The \glspl{acf} of $Q$ and $P$ are calculated in order to analyze the dynamics of the modes of interest as
\begin{align}
    \label{eq:mode_acf}
    C_Q(t) =& \left < Q_\lambda (t') Q_\lambda (t+t') \right >,
\end{align}
which can be modeled as the \gls{acf} of a \gls{dho}.
The \gls{dho} is driven by a stochastic force and has a natural frequency $\omega_0$ and a damping $\Gamma$.
The \gls{acf} of the \gls{dho} splits into an underdamped regime ($\omega_0 > \Gamma/2$) and an overdamped regime ($\omega_0 < \Gamma/2$).
In the underdamped regime the solution of the \gls{dho} is
\begin{align}
    \label{eq:acf_Q_underdamped}
    C^\text{DHO}_Q(t) = A \mathrm{e}^{-t/\tau} \left ( \cos{\omega_e t} + \frac{\Gamma}{2\omega_e}\sin{\omega_e t} \right ),
\end{align}
where $\omega_e = \sqrt{\omega_0^2 - \frac{\Gamma^2}{4}}$, the relaxation time is $\tau = 2 / \Gamma$, and $A$ is the amplitude \cite{FraSlaErhWah2021}.
In the overdamped limit the solution becomes the sum of two exponential decays as
\begin{align}
    \label{eq:acf_Q_overdamped}
    C^\text{DHO}_Q(t) = \frac{A}{\tau_\text{L} - \tau_\text{S}} \left(  \tau_\text{L} \text{e}^{-t / \tau_\text{L}}  - \tau_\text{S} \text{e}^{-t / \tau_\text{S}} \right )
\end{align}
where
\begin{align*}
    \tau_\text{S,L} = \frac{\tau}{1 \pm \sqrt{1-(\omega_0\tau)^2}}.
\end{align*}
Here, $\tau_\text{S}$ and $\tau_\text{L}$ denote the short and long timescales, respectively.
If the natural frequency approaches zero (e.g., for continuous phase transitions driven by a soft mode) we thus expect $\tau_\text{L} \to \infty$ and $\tau_\text{S} \to \tau/2$.
In this limit the resulting \gls{acf}, $C^\text{DHO}_Q(t)$, would only consist of a single exponential decay, with a decay time approaching infinity, which corresponds to the behavior seen in Brownian motion.

Similar expressions are obtained for the \gls{acf} of the phonon velocity, which is $C^\text{DHO}_P(t) = - \frac{\mathrm{d}^2 }{\mathrm{d} t^2} C^\text{DHO}_Q(t)$.
For the overdamped case it becomes
\begin{align*}
C^\text{DHO}_P(t) =\frac{A}{\tau_\text{L}-\tau_\text{S}} \left(  \frac{1}{\tau_\text{S}} \text{e}^{-t / \tau_\text{S}} - \frac{1}{\tau_\text{L}} \text{e}^{-t / \tau_\text{L}}  \right )
\end{align*}
The \glspl{acf} for $Q$ and $P$ are fitted simultaneously to the \gls{dho} model in order to extract $\omega_0$ and $\Gamma$.

% ====================
%% Additional things
%% ===================
\section{Data availability}

The \gls{dft} data and the \gls{mlp} models are provided in a Zenodo dataset \cite{zenodo_dataset}.

\section*{Author contributions}
E.\ F. carried out the analysis based on MD simulations and wrote the first draft of the manuscript.
P.\ R. constructed the 4th order \glspl{fcp}, carried out the \gls{scp} analysis with \textsc{hiphive} and \textsc{SSCHA} as well as some of the \gls{dft} calculations.
M.\ R. and P.\ E. carried out the \gls{dft} calculations and M.\ R. trained the \gls{nep} model.
F.\ E. analyzed the \gls{dho} expressions.
T.\ T. constructed an interface between \textsc{hiphive} and \textsc{alamode} and carried out the \gls{scp} calculations with \textsc{alamode}.
P.\ E. supervised the project.
All authors reviewed and edited the manuscript.

\section*{Competing interests}
The authors declare no competing financial or non-financial interest.

\begin{acknowledgments}
This work was funded by the Swedish Research Council (grant numbers 2018-06482, 2020-04935, 2021-05072), the Swedish Energy Agency (grant No. 45410-1), the Area of Advance Nano at Chalmers, and the Chalmers Initiative for Advancement of Neutron and Synchrotron Techniques.
T.\ T. was supported by JSPS KAKENHI Grant No. 21K03424.
The computations were enabled by resources provided by the National Academic Infrastructure for Supercomputing in Sweden (NAISS) at NSC, C3SE, and PDC partially funded by the Swedish Research Council through grant agreement no.\ 2022-06725.
\end{acknowledgments}


\begin{thebibliography}{10}
\expandafter\ifx\csname url\endcsname\relax
  \def\url#1{\texttt{#1}}\fi
\expandafter\ifx\csname urlprefix\endcsname\relax\def\urlprefix{URL }\fi
\providecommand{\bibinfo}[2]{#2}
\providecommand{\eprint}[2][]{\url{#2}}

\bibitem{Zim60}
\bibinfo{author}{Ziman, J.~M.}
\newblock \emph{\bibinfo{title}{Electrons and Phonons}}
  (\bibinfo{publisher}{Oxford University Press, London}, \bibinfo{year}{1960}).

\bibitem{SunAll2010}
\bibinfo{author}{Sun, T.} \& \bibinfo{author}{Allen, P.~B.}
\newblock \bibinfo{title}{Lattice thermal conductivity: Computations and theory
  of the high-temperature breakdown of the phonon-gas model}.
\newblock \emph{\bibinfo{journal}{Physical Review B}}
  \textbf{\bibinfo{volume}{82}}, \bibinfo{pages}{224305}
  (\bibinfo{year}{2010}).
\newblock \urlprefix\url{https://link.aps.org/doi/10.1103/PhysRevB.82.224305}.

\bibitem{SunSheAll2010}
\bibinfo{author}{Sun, T.}, \bibinfo{author}{Shen, X.} \&
  \bibinfo{author}{Allen, P.~B.}
\newblock \bibinfo{title}{Phonon quasiparticles and anharmonic perturbation
  theory tested by molecular dynamics on a model system}.
\newblock \emph{\bibinfo{journal}{Physical Review B}}
  \textbf{\bibinfo{volume}{82}}, \bibinfo{pages}{224304}
  (\bibinfo{year}{2010}).
\newblock \urlprefix\url{https://link.aps.org/doi/10.1103/PhysRevB.82.224304}.

\bibitem{ZhaSunWen2014}
\bibinfo{author}{Zhang, D.-B.}, \bibinfo{author}{Sun, T.} \&
  \bibinfo{author}{Wentzcovitch, R.~M.}
\newblock \bibinfo{title}{Phonon quasiparticles and anharmonic free energy in
  complex systems}.
\newblock \emph{\bibinfo{journal}{Physical Review Letters}}
  \textbf{\bibinfo{volume}{112}}, \bibinfo{pages}{058501}
  (\bibinfo{year}{2014}).
\newblock
  \urlprefix\url{https://link.aps.org/doi/10.1103/PhysRevLett.112.058501}.

\bibitem{Lv2016}
\bibinfo{author}{Lv, W.} \& \bibinfo{author}{Henry, A.}
\newblock \bibinfo{title}{Examining the validity of the phonon gas model in
  amorphous materials}.
\newblock \emph{\bibinfo{journal}{Scientific Reports}}
  \textbf{\bibinfo{volume}{6}}, \bibinfo{pages}{37675} (\bibinfo{year}{2016}).
\newblock \urlprefix\url{https://doi.org/10.1038/srep37675}.

\bibitem{Isaeva2019}
\bibinfo{author}{Isaeva, L.}, \bibinfo{author}{Barbalinardo, G.},
  \bibinfo{author}{Donadio, D.} \& \bibinfo{author}{Baroni, S.}
\newblock \bibinfo{title}{Modeling heat transport in crystals and glasses from
  a unified lattice-dynamical approach}.
\newblock \emph{\bibinfo{journal}{Nature Communications}}
  \textbf{\bibinfo{volume}{10}}, \bibinfo{pages}{3853} (\bibinfo{year}{2019}).

\bibitem{Silverman1974}
\bibinfo{author}{Silverman, B.~D.}
\newblock \bibinfo{title}{Collision-broadened phonon line shape in the
  overdamped or hydrodynamic regime}.
\newblock \emph{\bibinfo{journal}{Physical Review B}}
  \textbf{\bibinfo{volume}{9}}, \bibinfo{pages}{203--208}
  (\bibinfo{year}{1974}).
\newblock \urlprefix\url{https://link.aps.org/doi/10.1103/PhysRevB.9.203}.

\bibitem{SchSto1978}
\bibinfo{author}{Schneider, T.} \& \bibinfo{author}{Stoll, E.}
\newblock \bibinfo{title}{Molecular-dynamics study of a three-dimensional
  one-component model for distortive phase transitions}.
\newblock \emph{\bibinfo{journal}{Physical Review B}}
  \textbf{\bibinfo{volume}{17}}, \bibinfo{pages}{1302--1322}
  (\bibinfo{year}{1978}).
\newblock \urlprefix\url{https://link.aps.org/doi/10.1103/PhysRevB.17.1302}.

\bibitem{PetHeiTra91}
\bibinfo{author}{Petry, W.} \emph{et~al.}
\newblock \bibinfo{title}{Phonon dispersion of the bcc phase of group-{IV}
  metals. {I}. bcc titanium}.
\newblock \emph{\bibinfo{journal}{Physical Review B}}
  \textbf{\bibinfo{volume}{43}}, \bibinfo{pages}{10933--10947}
  (\bibinfo{year}{1991}).

\bibitem{FraErh20}
\bibinfo{author}{Fransson, E.} \& \bibinfo{author}{Erhart, P.}
\newblock \bibinfo{title}{Defects from phonons: Atomic transport by concerted
  motion in simple crystalline metals}.
\newblock \emph{\bibinfo{journal}{Acta Materialia}}
  \textbf{\bibinfo{volume}{196}}, \bibinfo{pages}{770} (\bibinfo{year}{2020}).

\bibitem{FraSlaErhWah2021}
\bibinfo{author}{Fransson, E.}, \bibinfo{author}{Slabanja, M.},
  \bibinfo{author}{Erhart, P.} \& \bibinfo{author}{Wahnstr\"{o}m, G.}
\newblock \bibinfo{title}{dynasor {\textemdash}a tool for extracting dynamical
  structure factors and current correlation functions from molecular dynamics
  simulations}.
\newblock \emph{\bibinfo{journal}{Advanced Theory and Simulations}}
  \textbf{\bibinfo{volume}{4}}, \bibinfo{pages}{2000240}
  (\bibinfo{year}{2021}).
\newblock \urlprefix\url{https://doi.org/10.1002/adts.202000240}.

\bibitem{Kim2021}
\bibinfo{author}{Kim, S.~E.} \emph{et~al.}
\newblock \bibinfo{title}{Extremely anisotropic van der waals thermal
  conductors}.
\newblock \emph{\bibinfo{journal}{Nature}} \textbf{\bibinfo{volume}{597}},
  \bibinfo{pages}{660--665} (\bibinfo{year}{2021}).
\newblock \urlprefix\url{https://doi.org/10.1038/s41586-021-03867-8}.

\bibitem{Nakamura1975}
\bibinfo{author}{Nakamura, T.}
\newblock \bibinfo{title}{Light scattering studies on soft phonon phase
  transitions}.
\newblock \emph{\bibinfo{journal}{Ferroelectrics}}
  \textbf{\bibinfo{volume}{9}}, \bibinfo{pages}{159--169}
  (\bibinfo{year}{1975}).

\bibitem{Nakamura1992}
\bibinfo{author}{Nakamura, T.}
\newblock \bibinfo{title}{Soft phonon in batio3}.
\newblock \emph{\bibinfo{journal}{Ferroelectrics}}
  \textbf{\bibinfo{volume}{137}}, \bibinfo{pages}{65--88}
  (\bibinfo{year}{1992}).

\bibitem{Dove1997}
\bibinfo{author}{Dove, M.~T.}
\newblock \bibinfo{title}{Theory of displacive phase transitions in minerals}.
\newblock \emph{\bibinfo{journal}{American Mineralogist}}
  \textbf{\bibinfo{volume}{82}}, \bibinfo{pages}{213--244}
  (\bibinfo{year}{1997}).
\newblock \urlprefix\url{https://doi.org/10.2138/am-1997-3-401}.

\bibitem{Ehsan2021}
\bibinfo{author}{Ehsan, S.}, \bibinfo{author}{Arrigoni, M.},
  \bibinfo{author}{Madsen, G. K.~H.}, \bibinfo{author}{Blaha, P.} \&
  \bibinfo{author}{Tr\"{o}ster, A.}
\newblock \bibinfo{title}{First-principles self-consistent phonon approach to
  the study of the vibrational properties and structural phase transition of
  ${\mathrm{batio}}_{3}$}.
\newblock \emph{\bibinfo{journal}{Physical Review B}}
  \textbf{\bibinfo{volume}{103}}, \bibinfo{pages}{094108}
  (\bibinfo{year}{2021}).

\bibitem{Verdi2022}
\bibinfo{author}{Verdi, C.}, \bibinfo{author}{Ranalli, L.},
  \bibinfo{author}{Franchini, C.} \& \bibinfo{author}{Kresse, G.}
\newblock \bibinfo{title}{Quantum paraelectricity and structural phase
  transitions in strontium titanate beyond density functional theory}.
\newblock \emph{\bibinfo{journal}{Phys. Rev. Mater.}}
  \textbf{\bibinfo{volume}{7}}, \bibinfo{pages}{L030801}
  (\bibinfo{year}{2023}).
\newblock
  \urlprefix\url{https://link.aps.org/doi/10.1103/PhysRevMaterials.7.L030801}.

\bibitem{Songvilay2019}
\bibinfo{author}{Songvilay, M.} \emph{et~al.}
\newblock \bibinfo{title}{Common acoustic phonon lifetimes in inorganic and
  hybrid lead halide perovskites}.
\newblock \emph{\bibinfo{journal}{Physical Review Materials}}
  \textbf{\bibinfo{volume}{3}}, \bibinfo{pages}{093602} (\bibinfo{year}{2019}).
\newblock
  \urlprefix\url{https://link.aps.org/doi/10.1103/PhysRevMaterials.3.093602}.

\bibitem{Lanigan2021}
\bibinfo{author}{Lanigan-Atkins, T.} \emph{et~al.}
\newblock \bibinfo{title}{Two-dimensional overdamped fluctuations of the soft
  perovskite lattice in {CsPbBr3}}.
\newblock \emph{\bibinfo{journal}{Nature Materials}}
  \textbf{\bibinfo{volume}{20}}, \bibinfo{pages}{977--983}
  (\bibinfo{year}{2021}).

\bibitem{Stoumpos2013}
\bibinfo{author}{Stoumpos, C.~C.} \emph{et~al.}
\newblock \bibinfo{title}{Crystal growth of the perovskite semiconductor
  {CsPbBr$_3$}: A new material for high-energy radiation detection}.
\newblock \emph{\bibinfo{journal}{Crystal Growth {\&} Design}}
  \textbf{\bibinfo{volume}{13}}, \bibinfo{pages}{2722--2727}
  (\bibinfo{year}{2013}).
\newblock \urlprefix\url{https://doi.org/10.1021/cg400645t}.

\bibitem{Hirotsu1974}
\bibinfo{author}{Hirotsu, S.}, \bibinfo{author}{Harada, J.},
  \bibinfo{author}{Iizumi, M.} \& \bibinfo{author}{Gesi, K.}
\newblock \bibinfo{title}{Structural phase transitions in cspbbr3}.
\newblock \emph{\bibinfo{journal}{Journal of the Physical Society of Japan}}
  \textbf{\bibinfo{volume}{37}}, \bibinfo{pages}{1393--1398}
  (\bibinfo{year}{1974}).

\bibitem{Sharma1991}
\bibinfo{author}{Sharma, S.}, \bibinfo{author}{Weiden, N.} \&
  \bibinfo{author}{Weiss, A.}
\newblock \bibinfo{title}{Phase transitions in {CsSnCl$_3$} and {CsPbBr$_3$} an
  {NMR} and {NQR} study}.
\newblock \emph{\bibinfo{journal}{Zeitschrift f\"{u}r Naturforschung A}}
  \textbf{\bibinfo{volume}{46}}, \bibinfo{pages}{329--336}
  (\bibinfo{year}{1991}).
\newblock \urlprefix\url{https://doi.org/10.1515/zna-1991-0406}.

\bibitem{Rodov2003}
\bibinfo{author}{Rodov{\'{a}}, M.}, \bibinfo{author}{Bro{\v{z}}ek, J.},
  \bibinfo{author}{Kn{\'{\i}}{\v{z}}ek, K.} \& \bibinfo{author}{Nitsch, K.}
\newblock \bibinfo{title}{Phase transitions in ternary caesium lead bromide}.
\newblock \emph{\bibinfo{journal}{Journal of Thermal Analysis and Calorimetry}}
  \textbf{\bibinfo{volume}{71}}, \bibinfo{pages}{667--673}
  (\bibinfo{year}{2003}).
\newblock \urlprefix\url{https://doi.org/10.1023/a:1022836800820}.

\bibitem{Lopez2020}
\bibinfo{author}{L{\'{o}}pez, C.~A.} \emph{et~al.}
\newblock \bibinfo{title}{Crystal structure features of {CsPbBr$_3$} perovskite
  prepared by mechanochemical synthesis}.
\newblock \emph{\bibinfo{journal}{{ACS} Omega}} \textbf{\bibinfo{volume}{5}},
  \bibinfo{pages}{5931--5938} (\bibinfo{year}{2020}).
\newblock \urlprefix\url{https://doi.org/10.1021/acsomega.9b04248}.

\bibitem{Malyshkin2020}
\bibinfo{author}{Malyshkin, D.} \emph{et~al.}
\newblock \bibinfo{title}{New phase transition in {CsPbBr}3}.
\newblock \emph{\bibinfo{journal}{Materials Letters}}
  \textbf{\bibinfo{volume}{278}}, \bibinfo{pages}{128458}
  (\bibinfo{year}{2020}).
\newblock \urlprefix\url{https://doi.org/10.1016/j.matlet.2020.128458}.

\bibitem{Huang2014}
\bibinfo{author}{Huang, L.-y.} \& \bibinfo{author}{Lambrecht, W. R.~L.}
\newblock \bibinfo{title}{Lattice dynamics in perovskite halides
  $\mathrm{CsSn}{X}_{3}$ with $x=\mathrm{I}, \mathrm{Br}, \mathrm{Cl}$}.
\newblock \emph{\bibinfo{journal}{Physical Review B}}
  \textbf{\bibinfo{volume}{90}}, \bibinfo{pages}{195201}
  (\bibinfo{year}{2014}).
\newblock \urlprefix\url{https://link.aps.org/doi/10.1103/PhysRevB.90.195201}.

\bibitem{daSilva2015}
\bibinfo{author}{da~Silva, E.~L.}, \bibinfo{author}{Skelton, J.~M.},
  \bibinfo{author}{Parker, S.~C.} \& \bibinfo{author}{Walsh, A.}
\newblock \bibinfo{title}{Phase stability and transformations in the halide
  perovskite ${\mathrm{cssni}}_{3}$}.
\newblock \emph{\bibinfo{journal}{Physical Review B}}
  \textbf{\bibinfo{volume}{91}}, \bibinfo{pages}{144107}
  (\bibinfo{year}{2015}).
\newblock \urlprefix\url{https://doi.org/10.1103/physrevb.91.144107}.

\bibitem{Yang2017}
\bibinfo{author}{Yang, R.~X.}, \bibinfo{author}{Skelton, J.~M.},
  \bibinfo{author}{da~Silva, E.~L.}, \bibinfo{author}{Frost, J.~M.} \&
  \bibinfo{author}{Walsh, A.}
\newblock \bibinfo{title}{Spontaneous octahedral tilting in the cubic inorganic
  cesium halide perovskites cssnx3 and cspbx3 (x = f, cl, br, i)}.
\newblock \emph{\bibinfo{journal}{The Journal of Physical Chemistry Letters}}
  \textbf{\bibinfo{volume}{8}}, \bibinfo{pages}{4720--4726}
  (\bibinfo{year}{2017}).
\newblock \urlprefix\url{https://doi.org/10.1021/acs.jpclett.7b02423}.

\bibitem{Klarbring2018}
\bibinfo{author}{Klarbring, J.} \& \bibinfo{author}{Simak, S.~I.}
\newblock \bibinfo{title}{Nature of the octahedral tilting phase transitions in
  perovskites: A case study of ${\mathrm{camno}}_{3}$}.
\newblock \emph{\bibinfo{journal}{Physical Review B}}
  \textbf{\bibinfo{volume}{97}}, \bibinfo{pages}{024108}
  (\bibinfo{year}{2018}).
\newblock \urlprefix\url{https://doi.org/10.1103/physrevb.97.024108}.

\bibitem{Yang2020}
\bibinfo{author}{Yang, R.~X.}, \bibinfo{author}{Skelton, J.~M.},
  \bibinfo{author}{da~Silva, E.~L.}, \bibinfo{author}{Frost, J.~M.} \&
  \bibinfo{author}{Walsh, A.}
\newblock \bibinfo{title}{Assessment of dynamic structural instabilities across
  24 cubic inorganic halide perovskites}.
\newblock \emph{\bibinfo{journal}{The Journal of Chemical Physics}}
  \textbf{\bibinfo{volume}{152}}, \bibinfo{pages}{024703}
  (\bibinfo{year}{2020}).
\newblock \urlprefix\url{https://doi.org/10.1063/1.5131575}.

\bibitem{Cohen2022}
\bibinfo{author}{Cohen, A.} \emph{et~al.}
\newblock \bibinfo{title}{Diverging expressions of anharmonicity in halide
  perovskites}.
\newblock \emph{\bibinfo{journal}{Advanced Materials}}
  \textbf{\bibinfo{volume}{34}}, \bibinfo{pages}{2107932}
  (\bibinfo{year}{2022}).
\newblock \urlprefix\url{https://doi.org/10.1002/adma.202107932}.

\bibitem{Klarbring2019}
\bibinfo{author}{Klarbring, J.}
\newblock \bibinfo{title}{Low-energy paths for octahedral tilting in inorganic
  halide perovskites}.
\newblock \emph{\bibinfo{journal}{Physical Review B}}
  \textbf{\bibinfo{volume}{99}}, \bibinfo{pages}{104105}
  (\bibinfo{year}{2019}).
\newblock \urlprefix\url{https://doi.org/10.1103/physrevb.99.104105}.

\bibitem{ZhuEgger2022}
\bibinfo{author}{Zhu, X.}, \bibinfo{author}{Caicedo-D{\'a}vila, S.},
  \bibinfo{author}{Gehrmann, C.} \& \bibinfo{author}{Egger, D.~A.}
\newblock \bibinfo{title}{Probing the disorder inside the cubic unit cell of
  halide perovskites from first-principles}.
\newblock \emph{\bibinfo{journal}{ACS Applied Materials \& Interfaces}}
  \textbf{\bibinfo{volume}{14}}, \bibinfo{pages}{22973--22981}
  (\bibinfo{year}{2022}).

\bibitem{Lahnsteiner2022}
\bibinfo{author}{Lahnsteiner, J.} \& \bibinfo{author}{Bokdam, M.}
\newblock \bibinfo{title}{Anharmonic lattice dynamics in large thermodynamic
  ensembles with machine-learning force fields:
  $\mathrm{Cs}\mathrm{Pb}\mathrm{Br}_{3}$, a phonon liquid with {Cs} rattlers}.
\newblock \emph{\bibinfo{journal}{Physical Review B}}
  \textbf{\bibinfo{volume}{105}}, \bibinfo{pages}{024302}
  (\bibinfo{year}{2022}).
\newblock \urlprefix\url{https://link.aps.org/doi/10.1103/PhysRevB.105.024302}.

\bibitem{TadWis2022}
\bibinfo{author}{Tadano, T.} \& \bibinfo{author}{Saidi, W.~A.}
\newblock \bibinfo{title}{First-principles phonon quasiparticle theory applied
  to a strongly anharmonic halide perovskite}.
\newblock \emph{\bibinfo{journal}{Physical Review Letters}}
  \textbf{\bibinfo{volume}{129}}, \bibinfo{pages}{185901}
  (\bibinfo{year}{2022}).

\bibitem{FanZenZha21}
\bibinfo{author}{Fan, Z.} \emph{et~al.}
\newblock \bibinfo{title}{Neuroevolution machine learning potentials: Combining
  high accuracy and low cost in atomistic simulations and application to heat
  transport}.
\newblock \emph{\bibinfo{journal}{Physical Review B}}
  \textbf{\bibinfo{volume}{104}}, \bibinfo{pages}{104309}
  (\bibinfo{year}{2021}).
\newblock \urlprefix\url{https://link.aps.org/doi/10.1103/PhysRevB.104.104309}.

\bibitem{FanWanYin22}
\bibinfo{author}{Fan, Z.} \emph{et~al.}
\newblock \bibinfo{title}{{GPUMD}: A package for constructing accurate
  machine-learned potentials and performing highly efficient atomistic
  simulations}.
\newblock \emph{\bibinfo{journal}{The Journal of Chemical Physics}}
  \textbf{\bibinfo{volume}{157}}, \bibinfo{pages}{114801}
  (\bibinfo{year}{2022}).
\newblock \urlprefix\url{https://doi.org/10.1063/5.0106617}.

\bibitem{zenodo_dataset}
\bibinfo{title}{Zenodo dataset}.
\newblock \bibinfo{howpublished}{\url{10.5281/zenodo.7313504}}
  (\bibinfo{year}{2022}).

\bibitem{KreHaf93}
\bibinfo{author}{Kresse, G.} \& \bibinfo{author}{Hafner, J.}
\newblock \bibinfo{title}{Ab initio molecular dynamics for liquid metals}.
\newblock \emph{\bibinfo{journal}{Physical Review B}}
  \textbf{\bibinfo{volume}{47}}, \bibinfo{pages}{558--561}
  (\bibinfo{year}{1993}).

\bibitem{Blo94}
\bibinfo{author}{Bl\"ochl, P.~E.}
\newblock \bibinfo{title}{Projector augmented-wave method}.
\newblock \emph{\bibinfo{journal}{Physical Review B}}
  \textbf{\bibinfo{volume}{50}}, \bibinfo{pages}{17953--17979}
  (\bibinfo{year}{1994}).

\bibitem{KreFur96}
\bibinfo{author}{Kresse, G.} \& \bibinfo{author}{Furthm\"uller, J.}
\newblock \bibinfo{title}{Efficiency of ab-initio total energy calculations for
  metals and semiconductors using a plane-wave basis set}.
\newblock \emph{\bibinfo{journal}{Computational Materials Science}}
  \textbf{\bibinfo{volume}{6}}, \bibinfo{pages}{15--50} (\bibinfo{year}{1996}).

\bibitem{SunRuzPer15}
\bibinfo{author}{Sun, J.}, \bibinfo{author}{Ruzsinszky, A.} \&
  \bibinfo{author}{Perdew, J.~P.}
\newblock \bibinfo{title}{{Strongly Constrained and Appropriately Normed
  Semilocal Density Functional}}.
\newblock \emph{\bibinfo{journal}{Physical Review Letters}}
  \textbf{\bibinfo{volume}{115}}, \bibinfo{pages}{036402}
  (\bibinfo{year}{2015}).

\bibitem{Larsen2017}
\bibinfo{author}{Larsen, A.~H.} \emph{et~al.}
\newblock \bibinfo{title}{The atomic simulation environment—a python library
  for working with atoms}.
\newblock \emph{\bibinfo{journal}{Journal of Physics: Condensed Matter}}
  \textbf{\bibinfo{volume}{29}}, \bibinfo{pages}{273002}
  (\bibinfo{year}{2017}).
\newblock \urlprefix\url{https://dx.doi.org/10.1088/1361-648X/aa680e}.

\bibitem{calorine}
\bibinfo{title}{\textsc{calorine}}.
\newblock
  \bibinfo{howpublished}{\url{https://gitlab.com/materials-modeling/calorine}}
  (\bibinfo{year}{2022}).
\newblock \bibinfo{note}{Accessed: 2023-02-10}.

\bibitem{Esfarjani2020}
\bibinfo{author}{Esfarjani, K.} \& \bibinfo{author}{Liang, Y.}
\newblock \bibinfo{title}{Thermodynamics of anharmonic lattices from first
  principles}.
\newblock In \emph{\bibinfo{booktitle}{Nanoscale Energy Transport}}, 2053-2563,
  \bibinfo{pages}{7--1 to 7--35} (\bibinfo{publisher}{IOP Publishing},
  \bibinfo{address}{Bristol England}, \bibinfo{year}{2020}).
\newblock \urlprefix\url{http://dx.doi.org/10.1088/978-0-7503-1738-2ch7}.

\bibitem{Kong2009}
\bibinfo{author}{Kong, L.~T.}, \bibinfo{author}{Bartels, G.},
  \bibinfo{author}{Campa{\~{n}}{\'{a}}, C.}, \bibinfo{author}{Denniston, C.} \&
  \bibinfo{author}{M\"{u}ser, M.~H.}
\newblock \bibinfo{title}{Implementation of green{\textquotesingle}s function
  molecular dynamics: An extension to {LAMMPS}}.
\newblock \emph{\bibinfo{journal}{Computer Physics Communications}}
  \textbf{\bibinfo{volume}{180}}, \bibinfo{pages}{1004--1010}
  (\bibinfo{year}{2009}).
\newblock \urlprefix\url{https://doi.org/10.1016/j.cpc.2008.12.035}.

\bibitem{Kong2011}
\bibinfo{author}{Kong, L.~T.}
\newblock \bibinfo{title}{Phonon dispersion measured directly from molecular
  dynamics simulations}.
\newblock \emph{\bibinfo{journal}{Computer Physics Communications}}
  \textbf{\bibinfo{volume}{182}}, \bibinfo{pages}{2201--2207}
  (\bibinfo{year}{2011}).
\newblock \urlprefix\url{https://doi.org/10.1016/j.cpc.2011.04.019}.

\bibitem{And12}
\bibinfo{author}{Andersson, T.}
\newblock \emph{\bibinfo{title}{One-shot free energy calculations for
  crystalline materials}}.
\newblock Master's thesis, \bibinfo{school}{Chalmers University of Technology}
  (\bibinfo{year}{2012}).

\bibitem{HelSteAbr13}
\bibinfo{author}{Hellman, O.}, \bibinfo{author}{Steneteg, P.},
  \bibinfo{author}{Abrikosov, I.~A.} \& \bibinfo{author}{Simak, S.~I.}
\newblock \bibinfo{title}{Temperature dependent effective potential method for
  accurate free energy calculations of solids}.
\newblock \emph{\bibinfo{journal}{Physical Review B}}
  \textbf{\bibinfo{volume}{87}}, \bibinfo{pages}{104111}
  (\bibinfo{year}{2013}).

\bibitem{EriFraErh19}
\bibinfo{author}{Eriksson, F.}, \bibinfo{author}{Fransson, E.} \&
  \bibinfo{author}{Erhart, P.}
\newblock \bibinfo{title}{The {{Hiphive Package}} for the {{Extraction}} of
  {{High}}-{{Order Force Constants}} by {{Machine Learning}}}.
\newblock \emph{\bibinfo{journal}{Advanced Theory and Simulations}}
  \textbf{\bibinfo{volume}{2}}, \bibinfo{pages}{1800184}
  (\bibinfo{year}{2019}).
\newblock
  \urlprefix\url{https://onlinelibrary.wiley.com/doi/abs/10.1002/adts.201800184}.

\bibitem{TadGohTsu14}
\bibinfo{author}{Tadano, T.}, \bibinfo{author}{Gohda, Y.} \&
  \bibinfo{author}{Tsuneyuki, S.}
\newblock \bibinfo{title}{Anharmonic force constants extracted from
  first-principles molecular dynamics: applications to heat transfer
  simulations}.
\newblock \emph{\bibinfo{journal}{Journal of Physics: Condensed Matter}}
  \textbf{\bibinfo{volume}{26}}, \bibinfo{pages}{225402}
  (\bibinfo{year}{2014}).

\bibitem{MonBiaChe21}
\bibinfo{author}{Monacelli, L.} \emph{et~al.}
\newblock \bibinfo{title}{The stochastic self-consistent harmonic
  approximation: calculating vibrational properties of materials with full
  quantum and anharmonic effects}.
\newblock \emph{\bibinfo{journal}{Journal of Physics: Condensed Matter}}
  \textbf{\bibinfo{volume}{33}}, \bibinfo{pages}{363001}
  (\bibinfo{year}{2021}).

\bibitem{Stukowski2010}
\bibinfo{author}{Stukowski, A.}
\newblock \bibinfo{title}{Visualization and analysis of atomistic simulation
  data with ovito–the open visualization tool}.
\newblock \emph{\bibinfo{journal}{Modelling and Simulation in Materials Science
  and Engineering}} \textbf{\bibinfo{volume}{18}}, \bibinfo{pages}{015012}
  (\bibinfo{year}{2009}).
\newblock \urlprefix\url{https://dx.doi.org/10.1088/0965-0393/18/1/015012}.

\bibitem{FraWikErh2023}
\bibinfo{author}{Fransson, E.}, \bibinfo{author}{Wiktor, J.} \&
  \bibinfo{author}{Erhart, P.}
\newblock \bibinfo{title}{Phase transitions in inorganic halide perovskites
  from machine learning potentials} (\bibinfo{year}{2023}).
\newblock \urlprefix\url{https://doi.org/10.48550/arxiv.2301.03497}.
\newblock \bibinfo{note}{ArXiv.2301.03497}.

\bibitem{Fabini2016}
\bibinfo{author}{Fabini, D.~H.} \emph{et~al.}
\newblock \bibinfo{title}{Dynamic stereochemical activity of the sn2+ lone pair
  in perovskite cssnbr3}.
\newblock \emph{\bibinfo{journal}{Journal of the American Chemical Society}}
  \textbf{\bibinfo{volume}{138}}, \bibinfo{pages}{11820--11832}
  (\bibinfo{year}{2016}).
\newblock \urlprefix\url{https://doi.org/10.1021/jacs.6b06287}.

\bibitem{Bertolotti2017}
\bibinfo{author}{Bertolotti, F.} \emph{et~al.}
\newblock \bibinfo{title}{Coherent nanotwins and dynamic disorder in cesium
  lead halide perovskite nanocrystals}.
\newblock \emph{\bibinfo{journal}{{ACS} Nano}} \textbf{\bibinfo{volume}{11}},
  \bibinfo{pages}{3819--3831} (\bibinfo{year}{2017}).
\newblock \urlprefix\url{https://doi.org/10.1021/acsnano.7b00017}.

\bibitem{Wiktor2017}
\bibinfo{author}{Wiktor, J.}, \bibinfo{author}{Rothlisberger, U.} \&
  \bibinfo{author}{Pasquarello, A.}
\newblock \bibinfo{title}{Predictive determination of band gaps of inorganic
  halide perovskites}.
\newblock \emph{\bibinfo{journal}{The Journal of Physical Chemistry Letters}}
  \textbf{\bibinfo{volume}{8}}, \bibinfo{pages}{5507--5512}
  (\bibinfo{year}{2017}).
\newblock \urlprefix\url{https://doi.org/10.1021/acs.jpclett.7b02648}.

\bibitem{Levin2021}
\bibinfo{author}{Levin, I.} \emph{et~al.}
\newblock \bibinfo{title}{Nanoscale-correlated octahedral rotations in
  ${\mathrm{bazro}}_{3}$}.
\newblock \emph{\bibinfo{journal}{Physical Review B}}
  \textbf{\bibinfo{volume}{104}} (\bibinfo{year}{2021}).
\newblock \urlprefix\url{https://doi.org/10.1103/physrevb.104.214109}.

\bibitem{Malavasi2022}
\bibinfo{author}{Malavasi, L.} \emph{et~al.}
\newblock \bibinfo{title}{Cubic or not cubic? the short-range order of tin
  halide perovskites} (\bibinfo{year}{2022}).
\newblock \urlprefix\url{https://doi.org/10.26434/chemrxiv-2022-3l933}.

\bibitem{Volpe2013}
\bibinfo{author}{Volpe, G.} \& \bibinfo{author}{Volpe, G.}
\newblock \bibinfo{title}{Simulation of a {Brownian} particle in an optical
  trap}.
\newblock \emph{\bibinfo{journal}{American Journal of Physics}}
  \textbf{\bibinfo{volume}{81}}, \bibinfo{pages}{224--230}
  (\bibinfo{year}{2013}).

\bibitem{Bianco2017}
\bibinfo{author}{Bianco, R.}, \bibinfo{author}{Errea, I.},
  \bibinfo{author}{Paulatto, L.}, \bibinfo{author}{Calandra, M.} \&
  \bibinfo{author}{Mauri, F.}
\newblock \bibinfo{title}{Second-order structural phase transitions, free
  energy curvature, and temperature-dependent anharmonic phonons in the
  self-consistent harmonic approximation: Theory and stochastic
  implementation}.
\newblock \emph{\bibinfo{journal}{Physical Review B}}
  \textbf{\bibinfo{volume}{96}}, \bibinfo{pages}{014111}
  (\bibinfo{year}{2017}).
\newblock \urlprefix\url{https://doi.org/10.1103/physrevb.96.014111}.

\bibitem{KorBelYan2018}
\bibinfo{author}{Korotaev, P.}, \bibinfo{author}{Belov, M.} \&
  \bibinfo{author}{Yanilkin, A.}
\newblock \bibinfo{title}{Reproducibility of vibrational free energy by
  different methods}.
\newblock \emph{\bibinfo{journal}{Computational Materials Science}}
  \textbf{\bibinfo{volume}{150}}, \bibinfo{pages}{47--53}
  (\bibinfo{year}{2018}).
\newblock
  \urlprefix\url{https://www.sciencedirect.com/science/article/pii/S0927025618302118}.

\bibitem{MetKli2019}
\bibinfo{author}{Metsanurk, E.} \& \bibinfo{author}{Klintenberg, M.}
\newblock \bibinfo{title}{Sampling-dependent systematic errors in effective
  harmonic models}.
\newblock \emph{\bibinfo{journal}{Physical Review B}}
  \textbf{\bibinfo{volume}{99}}, \bibinfo{pages}{184304}
  (\bibinfo{year}{2019}).
\newblock \urlprefix\url{https://link.aps.org/doi/10.1103/PhysRevB.99.184304}.

\bibitem{Tolborg2022}
\bibinfo{author}{Tolborg, K.} \& \bibinfo{author}{Walsh, A.}
\newblock \bibinfo{title}{Anharmonic and entropic stabilisation of cubic
  zirconia from first principles}.
\newblock \emph{\bibinfo{journal}{ChemRxiv}}  (\bibinfo{year}{2022}).
\newblock \urlprefix\url{https://doi.org/10.26434/chemrxiv-2022-lkzm9}.

\bibitem{Cochran1960}
\bibinfo{author}{Cochran, W.}
\newblock \bibinfo{title}{Crystal stability and the theory of
  ferroelectricity}.
\newblock \emph{\bibinfo{journal}{Advances in Physics}}
  \textbf{\bibinfo{volume}{9}}, \bibinfo{pages}{387--423}
  (\bibinfo{year}{1960}).
\newblock \urlprefix\url{https://doi.org/10.1080/00018736000101229}.

\bibitem{PytteFeder1969}
\bibinfo{author}{Pytte, E.} \& \bibinfo{author}{Feder, J.}
\newblock \bibinfo{title}{Theory of a structural phase transition in
  perovskite-type crystals}.
\newblock \emph{\bibinfo{journal}{Physical Review}}
  \textbf{\bibinfo{volume}{187}}, \bibinfo{pages}{1077--1088}
  (\bibinfo{year}{1969}).
\newblock \urlprefix\url{https://link.aps.org/doi/10.1103/PhysRev.187.1077}.

\bibitem{Scott1974}
\bibinfo{author}{Scott, J.~F.}
\newblock \bibinfo{title}{Soft-mode spectroscopy: Experimental studies of
  structural phase transitions}.
\newblock \emph{\bibinfo{journal}{Rev. Mod. Phys.}}
  \textbf{\bibinfo{volume}{46}}, \bibinfo{pages}{83--128}
  (\bibinfo{year}{1974}).
\newblock \urlprefix\url{https://link.aps.org/doi/10.1103/RevModPhys.46.83}.

\bibitem{Marronnier2018}
\bibinfo{author}{Marronnier, A.} \emph{et~al.}
\newblock \bibinfo{title}{Influence of disorder and anharmonic fluctuations on
  the dynamical rashba effect in purely inorganic lead-halide perovskites}.
\newblock \emph{\bibinfo{journal}{The Journal of Physical Chemistry C}}
  \textbf{\bibinfo{volume}{123}}, \bibinfo{pages}{291--298}
  (\bibinfo{year}{2018}).
\newblock \urlprefix\url{https://doi.org/10.1021/acs.jpcc.8b11288}.

\bibitem{Kirchartz2018}
\bibinfo{author}{Kirchartz, T.}, \bibinfo{author}{Markvart, T.},
  \bibinfo{author}{Rau, U.} \& \bibinfo{author}{Egger, D.~A.}
\newblock \bibinfo{title}{Impact of small phonon energies on the charge-carrier
  lifetimes in metal-halide perovskites}.
\newblock \emph{\bibinfo{journal}{The Journal of Physical Chemistry Letters}}
  \textbf{\bibinfo{volume}{9}}, \bibinfo{pages}{939--946}
  (\bibinfo{year}{2018}).
\newblock \urlprefix\url{https://doi.org/10.1021/acs.jpclett.7b03414}.

\bibitem{zhao2020polymorphous}
\bibinfo{author}{Zhao, X.-G.}, \bibinfo{author}{Dalpian, G.~M.},
  \bibinfo{author}{Wang, Z.} \& \bibinfo{author}{Zunger, A.}
\newblock \bibinfo{title}{Polymorphous nature of cubic halide perovskites}.
\newblock \emph{\bibinfo{journal}{Phys. Rev. B}}
  \textbf{\bibinfo{volume}{101}}, \bibinfo{pages}{155137}
  (\bibinfo{year}{2020}).
\newblock \urlprefix\url{https://link.aps.org/doi/10.1103/PhysRevB.101.155137}.

\bibitem{wang2022accurate}
\bibinfo{author}{Wang, H.}, \bibinfo{author}{Tal, A.},
  \bibinfo{author}{Bischoff, T.}, \bibinfo{author}{Gono, P.} \&
  \bibinfo{author}{Pasquarello, A.}
\newblock \bibinfo{title}{Accurate and efficient band-gap predictions for metal
  halide perovskites at finite temperature}.
\newblock \emph{\bibinfo{journal}{npj Computational Materials}}
  \textbf{\bibinfo{volume}{8}} (\bibinfo{year}{2022}).
\newblock \urlprefix\url{https://doi.org/10.1038/s41524-022-00869-6}.

\bibitem{CarTogTan2017}
\bibinfo{author}{Carreras, A.}, \bibinfo{author}{Togo, A.} \&
  \bibinfo{author}{Tanaka, I.}
\newblock \bibinfo{title}{Dynaphopy: A code for extracting phonon
  quasiparticles from molecular dynamics simulations}.
\newblock \emph{\bibinfo{journal}{Computer Physics Communications}}
  \textbf{\bibinfo{volume}{221}}, \bibinfo{pages}{221--234}
  (\bibinfo{year}{2017}).
\newblock
  \urlprefix\url{https://www.sciencedirect.com/science/article/pii/S0010465517302631}.

\bibitem{RohLiLuoHen2022}
\bibinfo{author}{Rohskopf, A.}, \bibinfo{author}{Li, R.}, \bibinfo{author}{Luo,
  T.} \& \bibinfo{author}{Henry, A.}
\newblock \bibinfo{title}{A computational framework for modeling and simulating
  vibrational mode dynamics}.
\newblock \emph{\bibinfo{journal}{Modelling and Simulation in Materials Science
  and Engineering}} \textbf{\bibinfo{volume}{30}}, \bibinfo{pages}{045010}
  (\bibinfo{year}{2022}).
\newblock \urlprefix\url{https://dx.doi.org/10.1088/1361-651X/ac5ebb}.

\bibitem{FanWeiVie2017}
\bibinfo{author}{Fan, Z.}, \bibinfo{author}{Chen, W.},
  \bibinfo{author}{Vierimaa, V.} \& \bibinfo{author}{Harju, A.}
\newblock \bibinfo{title}{Efficient molecular dynamics simulations with
  many-body potentials on graphics processing units}.
\newblock \emph{\bibinfo{journal}{Computer Physics Communications}}
  \textbf{\bibinfo{volume}{218}}, \bibinfo{pages}{10--16}
  (\bibinfo{year}{2017}).
\newblock \urlprefix\url{https://doi.org/10.1016/j.cpc.2017.05.003}.

\bibitem{TogTan15}
\bibinfo{author}{Togo, A.} \& \bibinfo{author}{Tanaka, I.}
\newblock \bibinfo{title}{First principles phonon calculations in materials
  science}.
\newblock \emph{\bibinfo{journal}{Scripta Materialia}}
  \textbf{\bibinfo{volume}{108}}, \bibinfo{pages}{1--5} (\bibinfo{year}{2015}).

\end{thebibliography}
\end{document}